\documentclass[twocolumn, superscriptaddress, prl, reprint, showpacs, citeautoscript]{revtex4}
\pdfoutput=1
\usepackage{graphicx}
\usepackage{amsmath}

\def\bra#1{\langle#1\vert}
\def\ket#1{\vert#1\rangle}
\def\vec#1{\mathbf{#1}}
\def\unit#1{\ensuremath{~\mathrm{#1}}}

\def\ext{\mathrm{ext}}
\def\eqn#1{Eqn.~(\ref{#1})}
\def\fig#1{Fig.~\ref{#1}}

\graphicspath{./}

\begin{document}

\title{Improving the conductance of carbon nanotube networks through resonant
 momentum exchange}

\author{Robert~A.~Bell}

\author{Michael~C.~Payne}

\affiliation{Theory of Condensed Matter Group, Cavendish Laboratory, Cambridge,
 UK} 

\author{Arash~A.~Mostofi}

\affiliation{Departments of Materials and Physics, and the Thomas Young Centre
 for Theory and Simulation of Materials, Imperial College London, London
 SW7 2AZ, UK}

\date{\today}

\begin{abstract}
We present a mechanism to improve the conductivity of carbon nanotube (CNT)  
networks by improving the conductance between CNTs of different chirality. We 
argue generally that a weak perturbation can greatly improve the inter-tube 
conductance by allowing momentum-conserving tunnelling. The mechanism is 
verified with a tight-binding model, allowing an investigation of its impact 
for a network containing a range of chiralities. We discuss practical 
implementations, and conclude that it may be effected by weak physical 
interactions, and therefore does not require chemical bonding to the CNTs.
\end{abstract}


\maketitle

The remarkable electronic properties of single-walled carbon nanotubes (CNTs) 
make them excellent candidates for electronic devices of the future 
\cite{Shulaker2013,Cao2013,Appenzeller2008,Behabtu2008,Cao2009}.

Whilst there exist nanoscale applications that employ CNTs individually 
\cite{Shulaker2013,Franklin2012}, considerable interest has focused on 
utilising bulk CNT networks, with applications including macroscopic 
lightweight wires \cite{Sundaram2011,Behabtu2013,Behabtu2008} and transparent 
conducting films \cite{Wu2004,Cao2009}.

Retaining the properties of individual CNTs when scaling up to the manufacture 
of these networks remains a significant challenge. For example, the 
conductivity of CNT wires is orders of magnitude lower than expected from the 
theoretical conductivity of individual CNTs 
\cite{Alvarenga2010,Sundaram2011,Behabtu2013,Behabtu2008}.

In a macroscopic CNT network, no single CNT spans the two ends; to traverse the 
device, electrons must travel through pathways involving many CNTs. The overall 
conductivity of a CNT network, therefore, is determined not only by the 
intrinsic (intra-tube) conductivity of the individual CNTs, but also by the 
morphology of the network \cite{Hu2004} and by the inter-tube conductivity. 

One route to improving network conductivity, therefore, is to improve 
inter-tube tunnelling.

There have been several studies on inter-tube transport between the ends of 
capped or terminated CNTs \cite{Ke2007,Qian2007,Martins2009}. The tendency for 
CNTs to align axially in bundles however\cite{Ge1994}, together with their high 
aspect ratio, gives a relatively larger side-wall contact region over which 
inter-tube tunnelling can be improved \cite{Li2011}.
Side-wall inter-tube conduction between perfect CNTs has been shown to be 
strongly suppressed for CNTs of different chirality when compared to those of 
the same chirality \cite{Maarouf2000,Tunney2006}. This is a consequence of the 
sensitivity of CNT electronic structure to chirality and the requirement for
both momentum and energy conservation in the tunnelling process.

As typical synthesis methods generate CNT networks with a range of chiralities, 
the overall conductance of the CNT network may be strongly suppressed even if 
the constituent CNTs are defect free, perfectly aligned and only metallic in 
character.

Improving side-wall conductance has proved difficult as modifications that 
strongly interact with the CNTs, say through covalently bonded groups
\cite{Lee2005}, defects\cite{Tunney2006}, or adsorbed transition metal 
ions\cite{Li2011}, dramatically reduce the intrinsic conductance of individual 
CNTs. 
Weak interactions with adsorbed molecules have been shown to improve side-wall 
tunnelling \cite{Mowbray2009}, but this mechanism requires molecular states 
close in energy to the CNT Fermi energy and, therefore, the effectiveness of 
this approach is dependent on extrinsic factors that may be difficult to 
control.

In this Letter we consider the general problem of electron transport between 
CNTs of arbitrary chirality. We focus on transport between metallic CNTs, but 
our conclusions equally apply to semiconducting ones and to bulk inter-sheet 
conduction in twisted bilayer graphene\cite{Ohta2012}.

Our main result is that transport between tubes of different chirality can be 
greatly enhanced by the presence of a \emph{weak} external potential that 
achieves momentum-conserving tunnelling by enabling momentum exchange.

The fact that the potential need only be weak removes the need to modify the 
bonding structure of the CNTs and, hence, the intrinsic CNT conductance is 
unaffected. Furthermore, for a potential that varies on length-scales larger 
than inter-atomic bond lengths, back-scattering is forbidden and the total 
conductivity of the network is only improved by its presence.
This potential is extrinsic to the CNTs themselves and could, for example, 
arise from physical interactions of the CNTs with their environment. 
Our analytical results are fully supported and verified by numerical 
tight-binding electron transport calculations.

\begin{figure}
   \centering
   \includegraphics[width=\columnwidth]{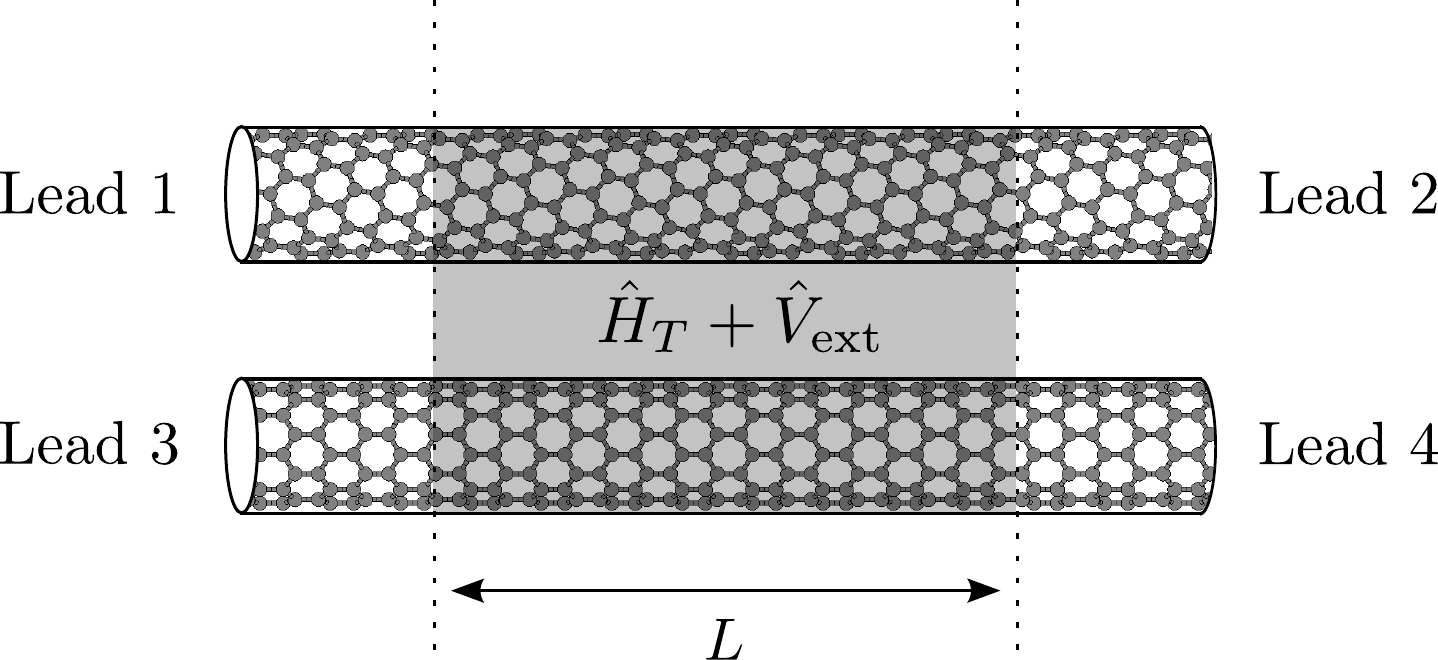}
   \caption{The four-lead side-wall tunnelling geometry considered in this
    work. The tunnelling and external interaction occur in the shaded region, 
    over a length-scale $L$, outside of which the leads are isolated.}
   \label{fig:intertube:setup}
\end{figure}

We consider a system containing two metallic CNTs of arbitrary chirality 
axially aligned as shown schematically in \fig{fig:intertube:setup}.

We denote the intrinsic tunnelling interaction between the CNTs by the operator 
$\hat{H}_{T}$.

Taking $\ket{\psi_1}$ and $\ket{\psi_2}$ to be electronic states at the Fermi 
energy $E_{\rm F}$ localised in each CNT, respectively, the contribution from 
these states to the small bias conductance $G_{1\rightarrow2}$ can be 
calculated using linear response theory. 
To first order,
\begin{align}
   G_{1\rightarrow 2} &= 2 e^2 \Gamma^{(1)}_{1\rightarrow2}\rho_1(E_{\rm F}) ,\\
   \Gamma^{(1)}_{1\rightarrow2} &= \frac{2\pi}{\hbar}\rho_2(E_{\rm F})\big| 
   \bra{\psi_{1}}{\hat{H}}_T\ket{\psi_{2}}\big|^2,
   \label{eq:intertube:perturbation_conductivity}
\end{align}
where $\Gamma^{(1)}_{1\rightarrow2}$ is the first-order tunnelling rate between 
$\ket{\psi_{1}}$ and $\ket{\psi_{2}}$, and $\rho_{1}(E_{\rm F})$ and 
$\rho_{2}(E_{\rm F})$ are the densities of states at the Fermi level for the 
two CNTs.

The most relevant term in \eqn{eq:intertube:perturbation_conductivity} is the 
matrix element and its dependence on the momenta of the initial and final 
states and, in turn, on the chiralities of the two CNTs. 

\begin{figure}
   \centering
   \includegraphics[width=\columnwidth]{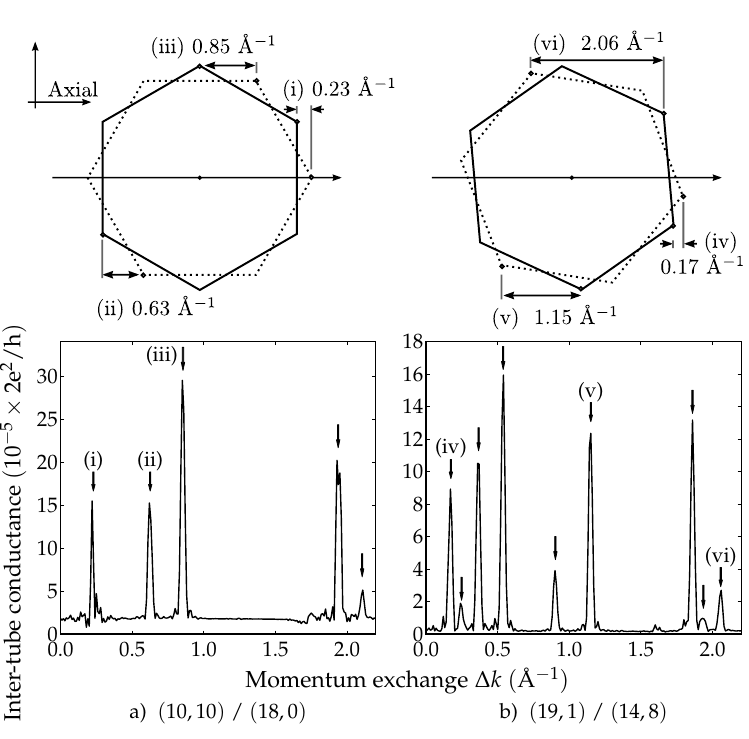}
   \caption{Top: schematic representation of tunnelling between CNTs. The 
   hexagons represent the first graphene Brillouin zone (BZ) for CNT pairs: a) 
   $(10,10)/(18,0)$ (dotted line/solid line) and b) $(19,1)/(14,8)$ (dotted/
   solid). Each BZ has been rotated so that the axial momentum lies along the 
   horizontal. The metallic states are at the BZ corners.
   Bottom: corresponding momentum exchange dependence of inter-tube conductance 
   at the Fermi level. Vertical arrows indicate the predicted momentum 
   exchanges corresponding to scattering between Dirac points. For selected 
   peaks, we show in the top panel the Dirac points involved in the scattering.}
   \label{fig:intertube:resonance_sweep} 
\end{figure}

For CNT diameters typical to networks, the electronic states of the CNTs and 
their momenta are well described by the band-structure of graphene under a zone-
folding approximation\cite{Charlier2007}. 
For two weakly-interacting axially-aligned CNTs we may consider two graphene 
Brillouin zones, each rotated by the chiral angle of the respective CNTs such 
that their axial momenta are aligned\cite{Maarouf2000}, as shown schematically 
in \fig{fig:intertube:resonance_sweep}.

Scattering between CNT states at the Fermi energy is equivalent to scattering 
between the Dirac points of these rotated Brillouin zones\footnote{In Ref.~
\onlinecite{Maarouf2000}, Eqns.~(2.11),~(2.12) apply to a general tunnelling 
interaction thus proving this statement.}, which in general have different 
momenta.
This momentum difference, containing both axial and azimuthal components, 
increases as the difference in chiral angle increases.

For two CNTs that are commensurate (i.e., the ratio of the unit cell lengths is 
a rational fraction), the tunnelling interaction is translationally invariant 
and momentum must be strictly conserved\cite{Maarouf2000}, resulting in the 
remarkable conclusion that conductance between pristine CNTs of different 
chirality is zero.

The more common scenario, however, is that the CNTs are incommensurate,
in which case translational invariance is broken \cite{Maarouf2000,Yoon2002}, 
and the requirement for momentum conservation is no longer strict. Nonetheless, 
it has been shown that inter-tube conductance at first order is still strongly 
suppressed for CNTs of different chirality due to the momentum mismatch of 
initial and final states \cite{Maarouf2000,Yoon2002}. 

Irrespective of whether commensurate or not, therefore, there is a strong 
suppression of inter-tube transport between CNTs of mismatched chirality.

How may the inter-tube conductance between CNTs of different chirality be 
improved? There are two general statements that can be made about any 
additional tunnelling interaction: (1) it must couple two states that are 
spatially separated and localised on different CNTs; (2) it must accept any 
momentum difference between the conducting states.

We consider the effect of an extrinsic \emph{weak} local potential 
$V_{\ext}(\vec{r})$ (which may be thought of as arising from the interaction of 
the environment with the CNTs) and consider the conditions it must satisfy to 
improve inter-tube conductance.

To improve conductance at first order, because the initial and final states are 
localised to different CNTs, $V_{\ext}$ would need to perturb the contact 
region between the CNTs where there is greatest spatial overlap.

Whilst defects\cite{Tunney2006}, functionalization\cite{Li2011} and structural 
deformations\cite{Tunney2006,Yoon2001} may achieve this, they may also 
significantly reduce the intrinsic intra-tube conductance
\cite{Lee2005,Choi2000} and, particularly in the case of side-wall 
functionalization, may increase the inter-tube separation and hence the 
tunnelling barrier.

In order to circumvent these problems, we consider alternative mechanisms and 
examine the second order scattering rate:
\begin{align}
   \Gamma^{(2)}_{1\rightarrow 2} &= \frac{2\pi}{\hbar} \rho_2(E_{\rm F}) 
   \Big| \sum_m \frac{\bra{\psi_1}{\hat{H}^{\prime}_{T}}\ket{m}\bra{m}
   {\hat{H}^{\prime}_{T}}\ket{\psi_2}}
   {E_{\rm F} - E_m} \Big|^2,
   \label{eqn:intertube:second_order_conductivity}
\end{align}
where $\hat{H}^{\prime}_{T}$ is the sum of the 
tunnelling interaction and the extrinsic potential, ${\hat{H}^{\prime}_{T} = 
\hat{H}_{T} + V_{\ext}}$, and the sum runs over all possible states $\ket{m}$. 
Expanding the matrix element gives four terms
\begin{align}
   \label{eqn:intertube:second_order_matrix_elements}
   \bra{\psi_1}&{\hat{H}^{\prime}_{T}}\ket{m}\bra{m}{\hat{H}^{\prime}_{T}}
   \ket{\psi_2} = \\
         &\bra{\psi_1}{\hat{H}_T}\ket{m}\bra{m}{\hat{H}_T}\ket{\psi_2} + 
         \bra{\psi_1}{\hat{H}_T}\ket{m}\bra{m}{V_{\ext}}\ket{\psi_2}~+
         \nonumber \\
         &\bra{\psi_1}{V_{\ext}}\ket{m}\bra{m}{\hat{H}_T}\ket{\psi_2} + 
         \bra{\psi_1}{V_{\ext}}\ket{m}\bra{m}{V_{\ext}}\ket{\psi_2}. \nonumber 
\end{align}
The first and final terms correspond to an even number of inter-tube 
scatterings and hence do not contribute to the inter-tube conductance.
The remaining two terms, however, can be large if $V_{\ext}$ strongly couples 
states in the same CNT but with different momenta. 

The third term, for example, represents scattering the initial state to an 
intermediate state $\ket{m}$ localised within the same CNT possibly with 
different momentum, which then scatters into the other CNT. If $V_{\ext}$ is 
chosen such that the intermediate state has the same momentum as the final 
state, then inter-tube tunnelling by $\hat{H}_T$ is no longer suppressed, and 
can be much larger than first order tunnelling. 

We validate the above mechanism with numerical results using a tight binding 
model. The system Hamiltonian is given by
\begin{equation}
   \hat{H} = \sum_{\alpha=1}^{2}\hat{H}_{\alpha} + \hat{H}_{T} + \hat{V}_{\ext},
\end{equation}
where $\alpha=1,2$ indexes the two CNTs.

The band structures of the CNTs are described using a $\pi$-orbital model 
including nearest neighbour hopping
\begin{equation}
   \hat{H}_{\alpha} = -t_0 \sum_{\left<i,j\right>}\big(c^{\dagger}_{i\alpha} 
   c_{j\alpha} 
   + c^{\dagger}_{j\alpha} c_{i\alpha}\big),
\end{equation}
where $c^{\dagger}_{i\alpha}$ and $c_{i\alpha}$ are the creation and 
annihilation operators, respectively, for an electron on atomic site $i$, at 
position $\vec{r}_i$, belonging to CNT $\alpha$. 

The summation is performed over nearest neighbour sites $\left<i,j\right>$ on a 
rolled hexagonal lattice, with interatomic bond lengths $a_{\rm C-C}=1.415
\unit{\AA}$. The Fermi energy is set to zero and the hopping parameter $t_0=2.77
\unit{eV}$.

The intrinsic tunnelling interaction between the CNTs is
\begin{equation}
   \hat{H}_T = -\sum_{ij} t_{ij} \big(c^{\dagger}_{i1} c_{j2} + 
   c^{\dagger}_{j2} c_{i1}\big),
\end{equation}
where atom sites $i$, $j$ are on different CNTs.
Following Refs.~\onlinecite{Maarouf2000} and \onlinecite{Tunney2006}, the 
tunnelling matrix element is
\begin{equation}
   t_{ij} = t_{\perp} e^{-|\vec{r}_i-\vec{r}_j|/\delta},
\end{equation}
with $t_{\perp}=492\unit{eV}$ and $\delta=0.5\unit{\AA}$. The separation 
between the surfaces of the CNTs is $3.4\unit{\AA}$.

For the external potential, we make the ansatz ${V_{\ext}(\vec{r}) = V_0 
\sin(\Delta k z)}$, (the $z$-axis is chosen as the CNT axial direction), which 
scatters strongly between states in the same CNT with axial momenta differing 
by $\Delta k$. 

\begin{figure*}
   \centering
   \includegraphics[width=0.9\textwidth]{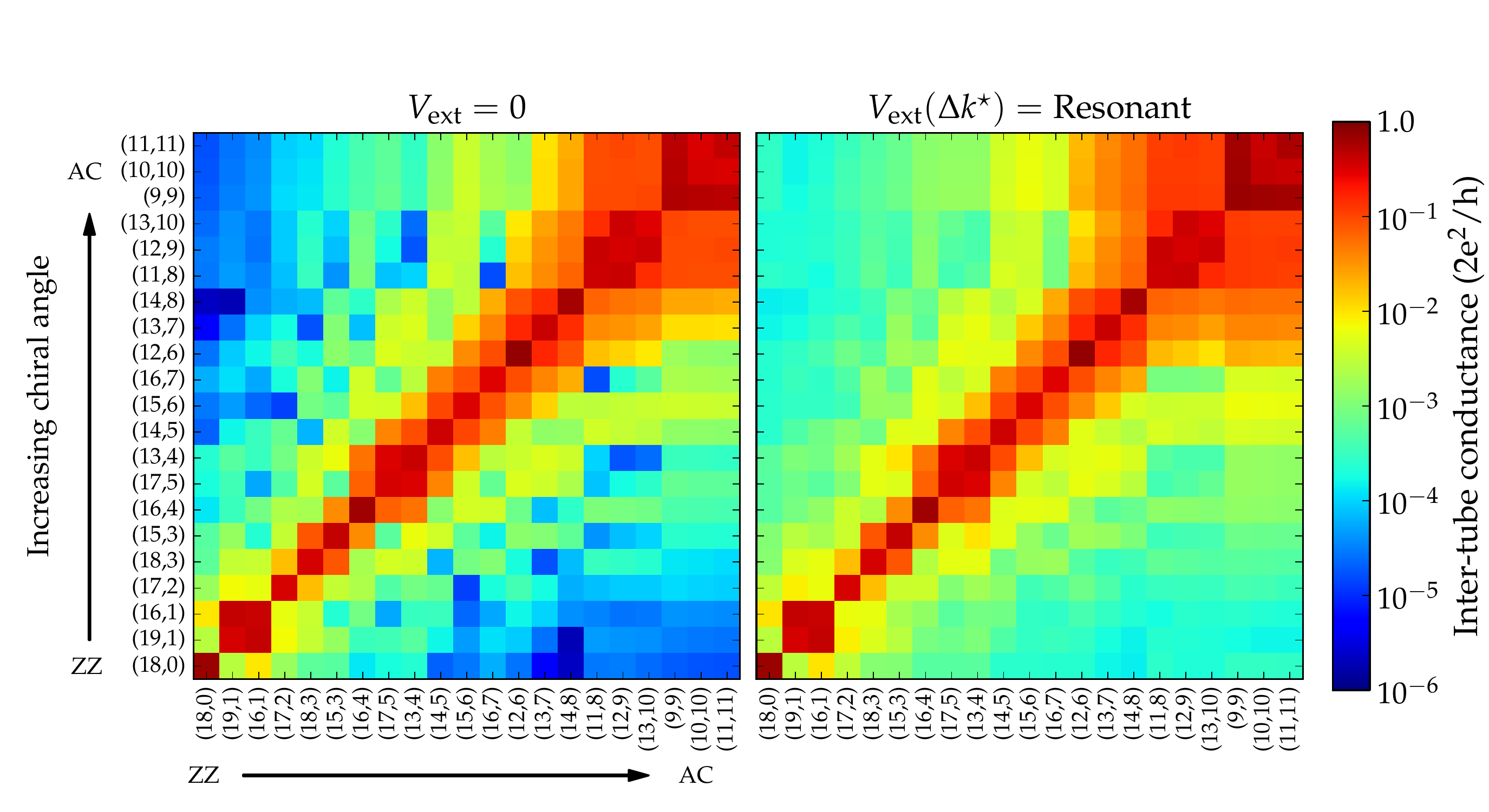}
   \caption{The inter-tube conductance between all pairs of metallic CNTs in 
   the diameter range $1.2-1.6\unit{nm}$. Left: intrinsic conductance 
   $V_{\mathrm{ext}} = 0$; right: maximum enhanced conductance 
   ($V_{\mathrm{ext}}$ resonant). The CNTs are ordered by increasing chiral 
   angle and labelled by the $(n,m)$ chiral indices.}
   \label{fig:intertube:array_improvement}
\end{figure*}

In the tight binding basis, this potential is approximated as diagonal
\begin{equation}
   \hat{V}_{\ext} = \sum_{i\alpha} V_0 \sin(\Delta k z_{i\alpha}) 
   c^{\dagger}_{i\alpha} c_{i\alpha},
\end{equation}
and therefore does not directly couple the two CNTs.
The amplitude $V_0=0.1\unit{eV}$ (i.e. ${V_0\ll t_0}$). A physical external 
potential can be considered as a sum of these Fourier components. We use a 
damping function to mutually isolate the leads with negligible contact-induced 
scattering\cite{Tunney2006}. Bulk inter-tube scattering occurs over a length 
$L=190\unit{\AA}$ (\fig{fig:intertube:setup}).

The ballistic conductance between the four semi-infinite leads is calculated 
using the Landauer-B\"{u}ttiker formalism via a Green's function approach
\cite{Datta1995}. We define the inter-tube conductance as the sum of the 
conductances between a lead and the two leads of the other CNT.

By ensuring that $V_{\ext}$ varies only on length-scales greater than the inter-
atomic bond length $a_{C-C}$, back-scatting is vanishing \cite{Ando1998}, 
corresponding to a maximum momentum exchange presented in this work 
$\Delta k_{\mathrm{max}}~=~\pi/a_{\mathrm{C-C}}$. We denote the resonant 
momentum exchanges and potentials as $\Delta k^{\star}$ and 
$V_{\ext}(\Delta k^{\star})$.

In \fig{fig:intertube:resonance_sweep} we plot for two CNT pairs the inter-tube 
conductance at the Fermi energy as a function of the momentum exchange 
$\Delta k$. The results are typical for CNT pairs with low intrinsic inter-tube 
conductance. We observe resonant peaks where the inter-tube conductance 
increases greatly. The momentum exchanges $\Delta k$ correspond precisely to 
the axial momentum difference between Dirac points. We indicate these predicted 
positions with arrows and show for selected peaks 
(\fig{fig:intertube:resonance_sweep}, top) the corresponding Dirac point 
scatterings.

The relative peak amplitudes are dependent on the form of the tight binding 
model and the spatial forms and energies of the intermediate states. Off 
momentum-resonance, the inter-tube conductance is unaffected by the weak 
potential, but for low momentum exchanges there is no decrease in the 
intra-tube conductance\cite{supplementary_material}.

A similar momentum resonance occurs between pairs of semiconductor CNTs
\cite{supplementary_material}.

Realistic CNT networks are compositionally disordered in the sense that they 
contain CNTs with a range of chiralities.

The left panel of \fig{fig:intertube:array_improvement} shows the intrinsic 
inter-tube conductance ($V_{\ext}=0$) at the Fermi energy between all pairs of 
metallic CNTs in the diameter range $1.2-1.6\unit{nm}$.

Where both CNTs are chiral, we show the conductance only for CNTs of opposite 
handedness; quantitatively equivalent results are obtained for pairs of CNTs of 
the same handedness\cite{supplementary_material}. The CNTs are ordered by 
increasing chiral angle. 

From the peak along the diagonal, it is evident that inter-tube conductance is 
large between CNTs of the same or similar chirality. As the difference in 
chirality between two CNTs increases, however, momentum-conserving tunnelling 
is not possible and inter-tube conductance is strongly suppressed.

The right panel of \fig{fig:intertube:array_improvement} shows the inter-tube 
conductance on the addition of the momentum-resonant potential 
$V_{\ext}(\Delta k^\star)$, at which the inter-tube conductance is a maximum.

The conductance between CNTs of the same chirality is unaffected and remains 
high, whilst there is a strong enhancement for CNTs of different chirality.

For different CNT pairs the inter-tube conductance is maximised by different 
momentum exchanges $\Delta k^{\star}$ (see, e.g., 
\fig{fig:intertube:resonance_sweep}), which are found to be distributed 
uniformly across a broad range of momentum \cite{supplementary_material}. The 
external potential, therefore, should contain a wide range of resonant momenta 
to improve the conductance of a network consisting of a wide range of CNT 
chiralities.

\begin{figure}
   \centering
   \includegraphics[width=\columnwidth]{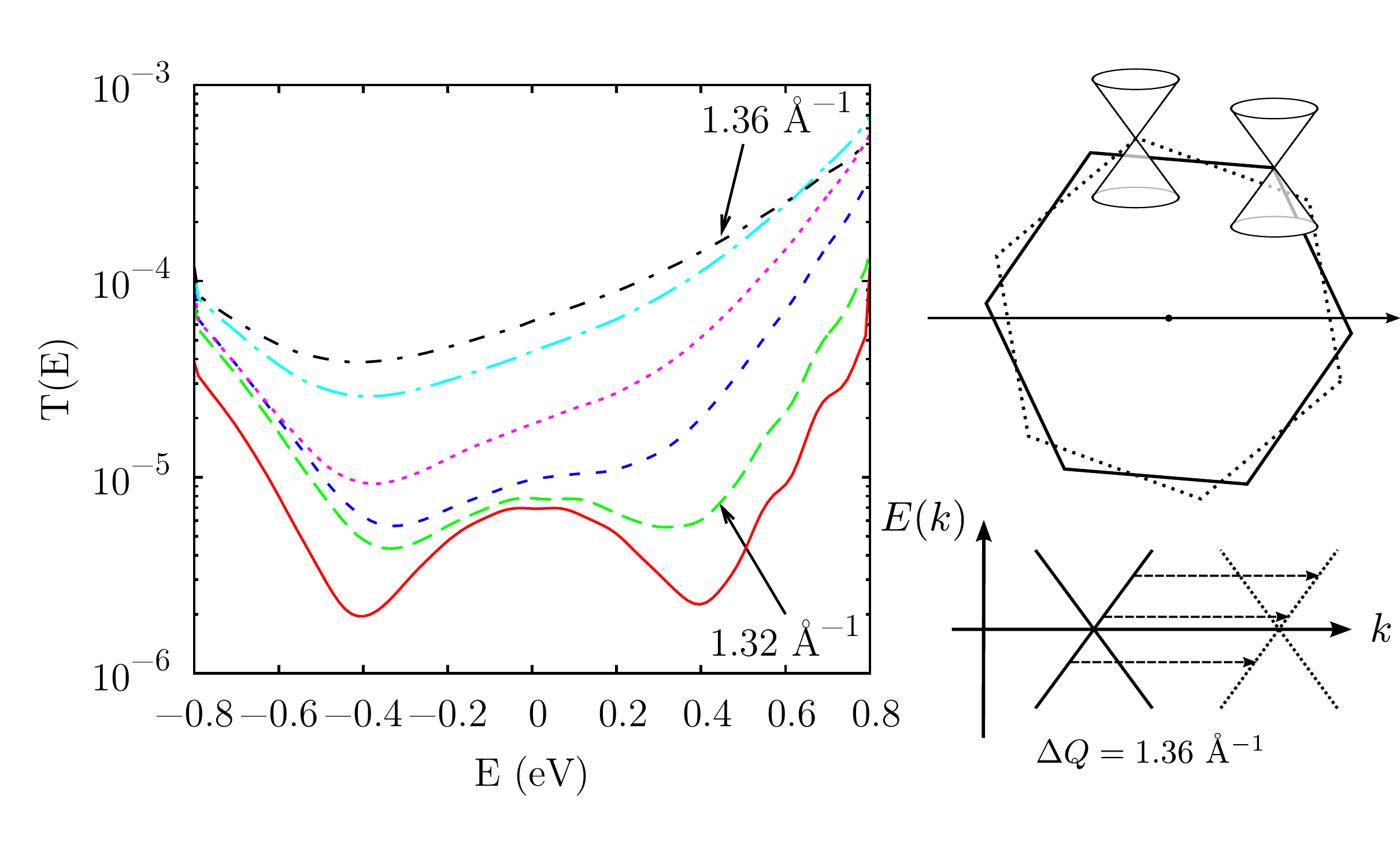}
   \caption{Left: the transmission between leads 1 and 4 for the 
   $(12,9)/(13,4)$ CNT pair as a function of electron energy for 
   $\hat{V}_{\ext}=0$ (solid red), and for momentum transfer increasing in 
   steps of $0.01\unit{\AA^{-1}}$ from $\Delta k=1.32\unit{\AA^{-1}}$ (dashed 
   green) to $\Delta k=1.36\unit{\AA^{-1}}$ (dot-dashed black) where momentum-
   resonance occurs. Transmission increases as resonance is approached. The 
   Fermi energy is zero. Right: the Dirac points involved in this resonance, 
   and the projection of the dispersion relation onto the axial momentum axis. 
   The momentum exchange, indicated by the arrows, is resonant over a wide 
   range of energies around the Fermi level.}
   \label{fig:intertube:energy_range}
\end{figure}

Finally, we consider this mechanism at energies away from the Fermi level. As a 
result of the linear dispersion relations around the Dirac points, for states 
travelling in the same direction the momentum exchange is resonant over a wide 
band of energies around the Fermi level. This is shown in 
\fig{fig:intertube:energy_range} where we plot the transmission spectrum 
between leads $1$ and $4$. (Transmission between leads $1$ and $3$ is resonant 
at the Fermi energy only\cite{supplementary_material}.) This mechanism is 
therefore insensitive to the small band gap in non-armchair metallic CNTs
\cite{Kane1997} and applies to doped CNT networks.

To summarise our results, we present a mechanism to improve the side-wall inter-
tube conductance between CNTs of different chirality. 
We find that whilst the best method is to reduce the range of chiralities 
present, significant improvements can be made with a weak external potential 
allowing momentum exchange with the environment, promoting momentum conserving 
scattering.

By ensuring that the potential varies slowly compared to the inter-atomic 
distance, back-scattering is vanishing\cite{Ando1998}, and as the mechanism is 
second order, the perturbation can be localised away from the contact region. 
Chemical bonding is not required, and instead weaker, long-range physical 
interactions can be utilised. Possible candidates include electrostatic 
influences of adsorbed molecules; substrate surface reconstructions; Moir\'{e} 
patterns induced in multi-walled CNTs \cite{Fukui2009}; phonon-assisted 
tunnelling; and polymer-wrapped CNTs\cite{Ma2008}. Indeed, enhancements due to 
this mechanism may have already been observed experimentally, and this work 
improves our understanding of, and mechanisms to improve, the conductance of 
CNT networks.

The authors thank S. Dubois for his implementation for calculating transmission 
spectra. R.A.B. acknowledges financial support from British Telecommunications 
and an EPSRC studentship. Calculations were performed using the the Darwin 
Supercomputer of the University of Cambridge High Performance Computing Service.

\bibliography{library}
\end{document}